# Ultra-compact modulators based on novel CMOS-compatible plasmonic materials

**Viktoriia E. Babicheva[1,2], Nathaniel Kinsey[2], Gururaj V. Naik[2], Andrei V. Lavrinenko[1], Vladimir M. Shalaev[2,3], Alexandra Boltasseva[2,1,\*]**

[1]*DTU Fotonik – Department of Photonics Engineering, Technical University of Denmark, Oersteds Plads 343, DK-2800 Kgs. Lyngby, Denmark*
[2]*School of Electrical & Computer Engineering and Birck Nanotechnology Center, Purdue University, 1205 West State Street, West Lafayette, IN 47907-2057 USA*
[3]*The Russian Quantum Center, Novaya Str., 100, BC "URAL", Skolkovo, Moscow region, 143025, Russia*
\* aeb@purdue.edu

**Abstract:** We propose several planar layouts of ultra-compact plasmonic waveguide modulators that utilize alternative CMOS-compatible materials. The modulation is efficiently achieved by tuning the carrier concentration in a transparent conducting oxide layer, thereby tuning the waveguide either in plasmonic resonance or off-resonance. Resonance significantly increases the absorption coefficient of the plasmonic waveguide, which enables larger modulation depth. We show that an extinction ratio of 86 dB/µm can be achieved, allowing for a 3-dB modulation depth in less than one micron at the telecommunication wavelength. Our multilayer structures can potentially be integrated with existing plasmonic and photonic waveguides as well as novel semiconductor-based hybrid photonic/electronic circuits.

## 1. Introduction

Plasmonics enables the merging between two major technologies: nanometer-scale electronics and ultra-fast photonics [1]. Metal-dielectric interfaces can support the waves known as surface plasmon polaritons (SPPs) that are tightly coupled to the interface, and allow manipulation of light at the nanoscale, overcoming the diffraction limit. Plasmonic technologies can lead to a new generation of fast, on-chip, nanoscale devices with unique capabilities [2,3]. To provide the basic nanophotonic circuitry functionalities, elementary plasmonic devices such as waveguides, modulators, sources, amplifiers, and photodetectors are required. Various designs of plasmonic waveguides have been proposed to achieve the highest mode localization and the lowest propagation losses [3]. In addition to waveguides, modulators are the most fundamental component for digital signal encoding and are paramount to the development of nanophotonic circuits. In this regard, opto-electronic modulators can be designed to achieve ultra-fast operational speeds in the 10's of GHz. Many plasmonic waveguide and modulator structures have been proposed and experimentally verified, but most of these structures use metals such as gold or silver, which are not CMOS compatible [4-22].

The promising development of chip-scale plasmonic devices with traditional noble metals is hindered by challenges such as high losses, continuous thin film growth, and non-tunable optical properties. Moreover, noble metals as plasmonic building blocks are not compatible with the established semiconductor manufacturing processes. This limits the ultimate applicability of such structures for future consumer devices. Recently, there have been efforts towards addressing the challenge of developing CMOS compatible material platforms for integrated plasmonic devices [23]. Similar to the advances in silicon technologies that led to the information revolution worldwide, the development of new CMOS compatible plasmonic materials with adjustable/tunable optical properties, could revolutionize the field of hybrid photonic/electronic devices. This technology would help to address the needs for faster, smaller and more efficient photonic systems, renewable energy, nanoscale fabrication, and biotechnologies. These new materials can bring exciting new functionalities that cannot be achieved with traditional metals.

Pioneering works in the search for new plasmonic materials [23,24] have suggested new intermediate carrier density materials such as transparent conducting oxides (TCOs) and transition metal nitrides as promising plasmonic building blocks with low loss, extraordinary tuning and modulation capabilities, and compatibility with standard semiconductor technology [25-32]. While many materials have been suggested as replacements for the traditional plasmonic metals, CMOS-compatible titanium nitride (TiN) is one of the best candidates [28,29]. Moreover, TiN is very thermally stable, bio-compatible, extremely hard (one of the hardest ceramics) and chemically stable; in particular, it does not oxidize like silver or copper. It was also shown that TiN provides higher mode confinement in comparison to gold [27]. This makes TiN a very promising material for telecommunication-range plasmonic waveguides.

One important advantage of TiN is that it can be grown epitaxially on many substrates including [100]-silicon, forming ultra-smooth and ultra-thin layers [29,33]. A final benefit of transition metal nitrides is that they are nonstoichiometric materials. Hence their optical properties depend greatly on the preparation conditions and can be varied based on the desired performance. In this study, we use the optical constants of TiN films optimized for plasmonic applications. The films were deposited at high temperature (800 °C) using reactive DC magnetron sputtering. This high temperature sputtering process is not utilized in the current semiconductor manufacturing processes for TiN deposition. Hence, optimization of the low temperature (less than 400 °C) deposition process used in CMOS industry is needed for TiN to suit plasmonic applications.

New intermediate carrier density materials offer the prospect of additional exotic properties beyond tailorable optical properties, lower losses and integration advantages [23]. TCOs can provide extraordinary tuning and modulation of their complex refractive indices, because their carrier concentrations can be changed over several orders of magnitude by applying an electric field [17,19,34]. Therefore, they are promising candidates for adding electro-optical capabilities to plasmonic devices [17,19]. In particular, a unity-order index change in a 5 nm thin Indium-Tin-Oxide (ITO) layer was demonstrated for a metal-insulator-metal (MIM) structure [34]. Tunability is accomplished by applying a bias, resulting in an electric field across the TCO layer. The resulting electric field causes a charge accumulation, or depletion, in the TCO layer (depending on the direction of electric field) which in turn changes the plasma frequency of the TCO, and consequently, its permittivity. The modulating speed is only RC limited and is expected to exceed 10's of GHz.

Several layouts of Si photonic modulators using a TCO as a dynamic layer were proposed [19,35]. The modulation can be characterized by the extinction ratio (ER), which shows how deep one can modulate a propagating wave per unit length of the absorption modulator. An extinction ratio of 1 dB/μm was demonstrated for a plasmonic modulator utilizing a metal-oxide-ITO stack on top of a silicon photonic waveguide [19]. Under an applied bias, the carrier concentration is changed from $6.8 \times 10^{20}$ to $1 \times 10^{19}$ cm$^{-3}$, and the propagation length is varied from 1.3 to 34 μm. However, because this structure uses a photonic mode, the miniaturization level of such a device is limited. Very high ER (up to 20 dB/μm) was achieved utilizing the epsilon-near-zero properties of Aluminum-Zinc-Oxide (AZO) [35]. Because of the small absolute value, a large portion of the field is localized within the layer and provides more efficient modulation.

ITO has also been implemented in an MIM waveguide structure to demonstrate a subwavelength plasmonic modulator [17], for which a five percent change in the average carrier density (from $9.25 \times 10^{20}$ to $9.7 \times 10^{20}$ cm$^{-3}$) was studied. The structure operates close to the surface plasmon resonance and an ER up to 2 dB/μm was theoretically predicted. The proposed design can be made extremely compact. However, due to the high confinement achievable in the MIM structure and the high losses associated with both the metal and ITO layers, the propagation length in this system was limited to approximately 0.5 μm.

ITO, and many other TCOs, may be deposited at relatively low temperatures (less than 300°C), which makes it possible to integrate them with standard silicon process despite the fact that they are not available in CMOS production lines. Similar techniques have been used to include lithium niobate crystals or electro-optic polymers on CMOS produced photonic chips [36]. Similar to other TCOs, ITO's properties depend strongly on fabrication conditions such as the annealing environment and temperature [24,28]. Since ITO provides a wide tunability of optical properties, the ITO layer optimization can be used to achieve better modulator performance (decreased losses in the structure) [37]. Periodic pattering of the active layer can provide further improvements allowing for propagation lengths of approximately 2.5 μm [37]. Ultra-compact designs can be achieved in a modulator layout based on an MIM waveguide with a very small (5 nm) gap [18]. Very small gap size leads to a very efficient change of the carrier concentration in a 2.5-nm-thick ITO layer. However, the practical implementation of the proposed device [18] is challenging.

In this paper, we will focus on developing active plasmonic devices that are compatible with CMOS processing techniques to allow for easy integration with current nanoelectronic devices (Fig. 1). We will suggest a variety of plasmonic modulator structures using transparent conducting oxides, which may serve as both the plasmonic material and as a dynamic element. In the design of the structures, both the device performance and its fabrication complexity are taken into account. We will numerically study these modulator geometries and compare their performance from different points of view. Modulation depth, propagation losses, mode size and their trade-off will be analyzed. Finally, we will investigate the integration of the modulator geometries with plasmonic waveguides, and compare their performance and fabrication complexity.

**2. Multilayer structures**

An example geometry of the proposed electro-optic plasmonic modulator integrated with stripe waveguides is shown in Fig. 1, along with the symmetric, long ranging SPP (LR-SPP) excitation mode, its direction of propagation through the structure, and the applied voltage for modulation. The waveguide sections are composed of a lower dielectric cladding (Fig. 1 shown in red) with a metal strip (grey) and a top dielectric cladding (red). This configuration allows for the LR-SPP mode to be utilized for low loss connections to and from the modulator. The modulator section is composed of a lower dielectric layer (Fig. 1 shown in red), a TCO layer used as both a plasmonic layer and an electrode (light blue), a dielectric layer to provide electrical insulation (light orange), and a top electrode (pink). However, we investigate two other modulator configurations in this paper, which will be discussed in the subsequent sections.

This proposed geometry provides several benefits when compared to other potential structures. The utilization of an electrical control system allows for the device to be easily integrated into traditional CMOS control circuitry. Also, due to the plasmonic nature of the modulator, the device can achieve a very small footprint that cannot be realized in traditional photonic elements. Secondly, the multilayer structure which forms the modulation, can achieve an extremely high absorption coefficient when this layer is modulated into resonance by an applied field. Therefore, a very small length is required to achieve 3-dB modulation of the signal. Due to these advantages, the proposed modulator configuration shows great promise for a CMOS compatible, on-chip electro-optic modulator.

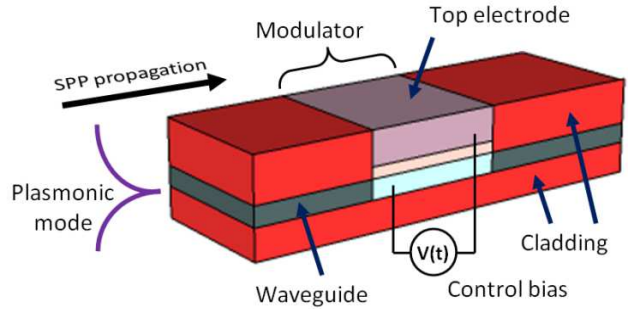

Fig. 1. General scheme of an ultra- compact modulator integrated with plasmonic waveguides. In this geometry, a stripe waveguide (grey) is used to bring a long ranging SPP mode to the modulator structure. Inside the modulator structure, the waveguide is replaced by a TCO (light blue) which is plasmonic at the wavelength of interest. A modulating voltage is applied between the TCO layer and a top electrode, which are separated by a thin dielectric spacer. This voltage alters the carrier concentration in the TCO resulting in change in the absorption which modulates the SPP. A second strip waveguide is used to propagate the modulated signal to the next component.

Due to obvious fabrication and integration advantages we consider modulators based on the strip waveguide geometry, as discussed in the previous section [6,7]. Stripe waveguides have low loss but have poor mode localization [6]. However, the relatively simple planar fabrication process provides an advantage to utilize the structure in realistic devices. Here we consider only the one-dimensional planar layout as an estimate of a stripe waveguide. In realistic structures, a finite-wide stripe waveguide will be used, and the propagation losses will depend on the geometrical parameters of the stripe. However, the performance is only marginally different from the one-dimensional structure, and the main dispersive features will be captured [5,38]. Therefore, for the purposes of this paper, the estimates provided by assuming a one-dimensional structure are suitable for the comparison of the various modulator geometries suggested.

The main element of the structure is the TCO layer. Since the TCO can possess plasmonic properties at telecom range, a thin TCO layer can guide the plasmonic mode as well as control the signal propagation. TCOs such as ITO, Gallium Zinc Oxide (GZO), and AZO have very similar properties and allow for an efficient change of carrier concentration. We chose GZO as it has shown the ability to achieve the highest plasma frequency of the three [39]. The permittivity of the GZO layer was taken from [39] and a carrier concentration in the GZO was determined using a Drude-Lorentz model fitting: $N_0 = 9.426 \times 10^{20}$ cm$^{-3}$. To estimate the modulator performance, we assume that under an applied voltage, the carrier concentration can be either decreased or increased by a factor of 2 ($N = 0.5N_0 \ldots 2N_0$). The calculated permittivity of GZO for these carrier concentrations is shown in Fig. 2a.

Including a TiN layer increases the mode localization, which influences the modulator's performance. Moreover, the TiN layer can also serve as a second electrode to apply bias to the GZO active layer. The permittivity of TiN is taken as experimentally measured $\varepsilon_{TiN} = -83.3 + 21.3i$ at $\lambda = 1.55$ μm (Fig. 2b). The TiN film was deposited at 800°C and the optical properties of 20 nm thin film was measured using spectroscopic ellipsometer (J.A. Woollam Co). High deposition temperature poses some fabrication and integration restrictions which must be taken into account. The materials beneath the TiN layer must withstand the TiN-deposition and etch conditions without degradation. Since the properties of the TCO degrade at high temperatures, the TCO layer must be deposited only after the deposition and patterning of the TiN layer.

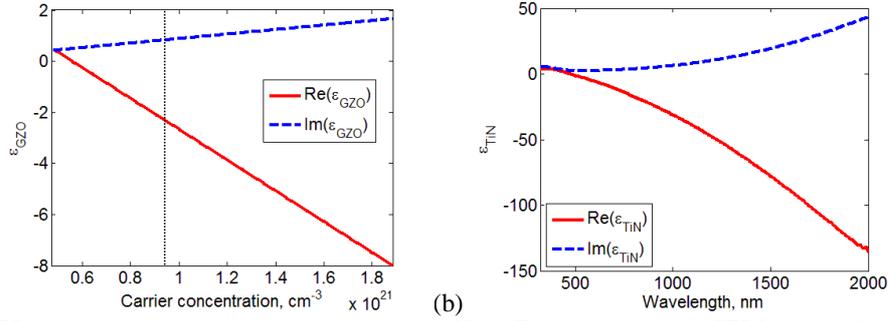

Fig. 2. (a) GZO permittivity versus its carrier concentration, $\lambda = 1.55$ μm. The permittivity of the GZO layer was taken from [39] and a carrier concentration in the GZO was determined using a Drude-Lorentz model fitting: $N_0 = 9.426 \times 10^{20}$ cm$^{-3}$. Black dotted line shows initial carrier concentration $N_0$ of as deposited GZO films. A factor of 2 change in the carrier concentration of 10 nm thin GZO film is assumed upon field-effect. (b) TiN permittivity extracted from spectroscopic ellipsometry measurements.

The simplest structure is shown in Fig. 3i-a where modulation is achieved by applying a bias across the GZO layer. For this structure, the zinc oxide (ZnO) layer serves as one electrode while the GZO film itself is the second. A dielectric layer between the TCO film and second conductive layer is required to provide electrical insulation. This layer should be made a thin as possible to reduce the voltage required to modulate the GZO carrier concentration.

For the design in Fig. 3 i-a, a thick $Si_3N_4$ layer can be utilized deposited, for example, on a silicon substrate. It is also preferable to have materials with similar indices on the top and bottom of the plasmonic layer. In this case, the conditions are similar to those required for long-range SPP mode propagation, and the mode losses are lower [6].

Furthermore, we studied designs that include TiN layers (Fig. 3b,c). The addition of the layer allows for the modulator to be easily integrated with external strip waveguides. It also provides tighter field confinement, which will result in a larger attenuation of the signal during modulation. Both layouts with a thick and thin TiN layer are studied. On the structures shown in Fig. 3, the central layers (TiN, silicon nitride insulation, and GZO) remain in the same configuration.

In this paper, we consider two main groups of the devices, one with low-index cladding (Fig. 3i) and another with high-index cladding (Fig. 3ii). Further in the text, ZnO, LP-CVD $Si_3N_4$ and PE-CVD silicon nitride (denoted in subsequent text by SiN), will be referred to as low-index materials. As we are interested in operation at the telecom wavelength of $\lambda = 1.55$ μm, the refractive indices used in the calculations are the following: $n_{ZnO} = 1.93$ [40], $n_{SiN} = 1.76$ (experimental characterization of samples after PE-CVD process) and $n_{Si3N4} = 1.97$ (after LP-CVD process). It should be mentioned that LP-CVD $Si_3N_4$ requires high temperature deposition which may degrade the properties of TCO layer. Hence, only PE-CVD SiN can be deposited after the TCO layer.

Furthermore, a high-index material can be utilized as a cladding. In this work, we consider silicon as a high-index cladding $n_{Si} = 3.48$ [41]. While amorphous silicon would be deposited as the upper cladding layer, for this investigation, we consider crystalline silicon and amorphous silicon to be identical in their optical properties at $\lambda=1.55$ μm. In particular, the next three structures (Fig. 3ii) are similar to the first three but with silicon layers as a top and bottom cladding. In these cases, either the silicon or TiN layer can be used as a second electrode.

## 3. Performance of the modulators

With all these considerations, six basic geometries were chosen as templates for modulator designs, operating at the telecom wavelength of $\lambda = 1.55$ μm. The dispersion equation was solved for the one-dimensional multilayer structures with varying carrier concentrations in the GZO. The thickness of all thin layers (GZO, TiN, SiN, $Si_3N_4$) is 10 nm. The top and bottom layers are assumed to be infinitely thick. Thus, we will refer to structure in Fig. 3a as "thick TiN" and Fig. 3b as "thin TiN". The structures in Fig. 3c are referred to as "without TiN".

At a particular carrier concentration, the GZO permittivity value satisfies the condition to achieve a plasmonic resonance in the multilayer structures, resulting in a significant increase in the absorption coefficient $\alpha_{max}$ (Fig. 3d). Therefore, the absorption coefficient $\alpha$, in the waveguide structure, strongly depends on carrier concentration N: $\alpha = \alpha(N)$. $\alpha_{max}$ is lower for structures without TiN, higher for a thick TiN layer and the highest with a thin TiN film (Table 1).

Modes of the structures with a GZO layer only (in Fig. 3a) have a quasi-symmetric electric field distribution. Adding a TiN layer increases the absorption coefficient of the multilayer structure because of the ohmic losses. Moreover, the structure with a thin TiN layer and low-index cladding (Fig. 3 i-c) supports only the mode with a quasi-asymmetric electric field profile, while the structure with the high-index cladding (Fig. 3 ii-c) supports both quasi-symmetric and quasi-asymmetric. In the quasi-asymmetric mode, the field is mostly localized near the waveguide (in contrast to quasi-symmetric mode, where the field is mostly spread outside the waveguide) such that the losses are even higher [42]. In this case, $α_{max}$ reaches 28 and 132 dB/μm for the low- and high-index cladding, respectively.

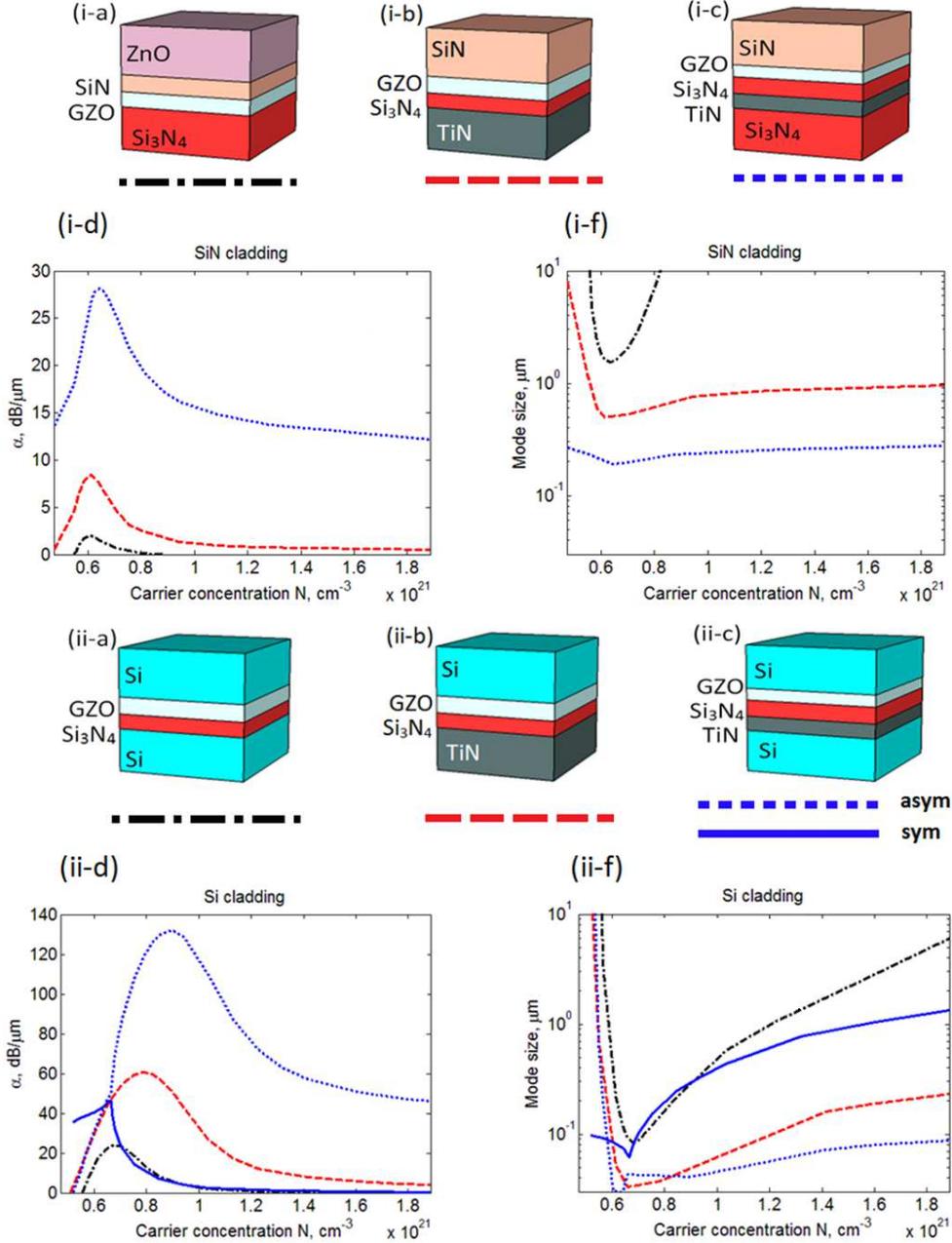

Fig. 3. Multilayer structures. Waveguide layers (GZO and TiN) are sandwiched or covered with (i): low-index materials, (ii): silicon as a high-index material. Further in the text, we will refer to designs (a) as "without TiN", (b) as "thick TiN", (c) as "thin TiN". (d) Absorption coefficients α for various carrier concentrations of GZO. Structures with high-index cladding (ii) show much higher absorption than structures with silicon nitride cladding (i). Notation "sym" and "asym" correspond to quasi-symmetric and quasi-asymmetric SPP modes respectively. (f) Mode size of the modulator structures versus carrier concentration in the GZO film. The absorption maximum is accompanied by highest mode localization. At lower carrier concentrations of GZO, modes are more spread-out because of the smaller magnitude of real permittivity of GZO.

Furthermore, we analyzed the size of the mode in all the proposed structures. In the case of a single interface, the 1/e point of the electric field corresponds to an 86% localization of electrical energy (1-$e^{-2}$ portion). To estimate the mode size of our multilayer structures, which have a complicated field profile, we define the mode size such that 86% of electrical energy is localized within the region (Fig. 4). Similar to other long-range SPP based waveguides, the structure suffers from low mode localization. It can be seen from Fig. 3f that structures without TiN have lower mode localization in comparison to those with TiN. Moreover, utilizing a high-index cladding significantly decreases mode size. In all cases, a decrease of α at lower N is accompanied by significant increase of mode size.

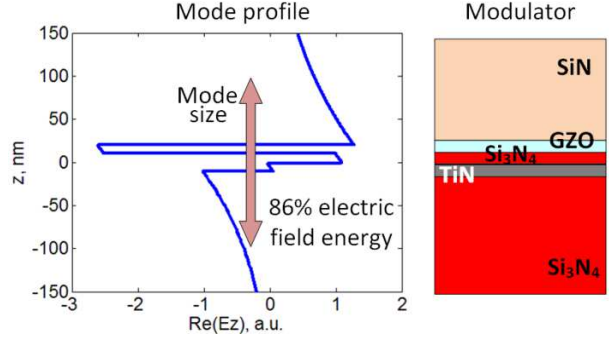

Fig. 4. Depiction of the mode profile for the geometry Fig. 3 i-c showing the definition of mode size. Due to the complexity of the structure and high concentration of electrical energy in the GZO layer, the traditional definition of mode is cannot be utilized as this would simply define the mode size as the thickness of the GZO layer. Here we define mode size as the distance which encompasses 86% of the electric field energy, a condition similar to that of the 1/e definition for single interface waveguide.

At some particular carrier concentration in GZO, the absorption coefficient reaches a maximum value $\alpha_{max}$. The extinction ratio ER, or modulation depth, can be defined as

$$ER = 8.68\,(\alpha_{max} - \alpha_{min}), \qquad (1)$$

where $\alpha_{min}$ is propagation loss in the transmittive state. Here we do not specify which states correspond to voltage on and voltage off states. It depends on how GZO layers are deposited and which carrier concentration is chosen as an initial value.

Either increasing or decreasing N from $N_0$ results in modulation of α. However, because of the large mode extension, or delocalization, at lower N, it is more preferable to operate at higher N (see Fig. 3f for small carrier concentrations). Thus, $\alpha_{min}$ is defined by

$$\alpha_{min} = \alpha(2N_0). \qquad (2)$$

Eq. (2) is valid for all the proposed structures apart from $Si_3N_4$/GZO/SiN/ZnO (Fig. 3 i-a). This is because with the low-index cladding and absence of TiN, a localized mode exists only for a narrow range of $N = 5 \times 10^{20}$ to $8 \times 10^{20}$ cm$^{-3}$. Modes larger than 10 μm are considered delocalized and the corresponding values for N will not be used in subsequent calculations. Thus, for this layout, $\alpha_{min} = \alpha\,(N = 8 \times 10^{20}$ cm$^{-3})$ is defined. While operation in this narrow range of N can be more preferable as it does not require a large change of N, it provides less tolerance to fabrication or design imperfections.

Similar to the value of $\alpha_{max}$, the ER is lower for the structures without TiN, higher for a thick TiN layer and the highest with a thin TiN film (see Table 1 for a comparison of values). The ER is 1.8-16 dB/μm for a silicon nitride cladding, and 24-86 dB/μm for a silicon cladding. In the latter case, less than a 35-nm-length active section is required to achieve 3 dB modulation.

A figure of merit FoM for such multilayer modulator structures can be defined as

$$FoM = ER/\alpha_{min}. \qquad (3)$$

It reflects a trade-off between the modulation depth and the loss of the signal in the transmitive state ($\alpha_{min}$). While the structures with a thin TiN layer provide the strongest resonance, α is also relatively high at large N. Such structures give the lowest performance.

The highest FoM is provided by structures without TiN. However, the lowest absorption in the transmitive state is accompanied by lowest mode localization (up to 10 μm). In Table 1 we summarize the ranges of mode extensions. In most cases, the minimum value corresponds to plasmonic resonance and the maximum to a carrier concentration $N = 2N_0$ (structure on Fig. 3i-a is an exception).

Calculations show that the high-index cladding designs possess the highest performance. Working with a 2x change in the carrier concentration of GZO, the studied plasmonic modulator can outperform previously proposed designs. For example, deeply subwavelength MIM structures were analyzed and a corresponding ER up to 12 dB/μm was theoretically predicted [17,18]. However, such high values are accompanied by high losses in transmittive state. The ratio of the absorption coefficients in the two states, FoM, is on order of 1. In our case, because of the possibility to detune from the plasmonic resonance, the absorption coefficient in the transmittive state can be relatively low. Utilizing a high index cladding makes the resonance more pronounced and the required change of carrier concentration is smaller. This structure also achieves transmittive state losses down to 0.06 dB/μm, which produced the highest FoM = 400. Thus, the Fig. 3ii-a geometry with high-index silicon claddings and without TiN, provides the highest performance for an ultra-compact plasmonic modulator.

Table 1. Summary of the characteristics of different structures. Performance comparison for planar modulator designs. Designs utilizing high-index materials show the highest performance.

| Structure (layers bottom to top) | $N(\alpha_{max})$, $10^{20}$cm$^{-3}$ | $\alpha_{max}$, dB/μm | $\alpha_{min}$, dB/μm | ER, dB/μm | FoM | Mode size, μm |
|---|---|---|---|---|---|---|
| Si$_3$N$_4$/GZO/SiN/ZnO (Fig. 3i-a) | 6.13 | 1.95 | 0.114 | 1.8 | 16 | 1.6 – 10 |
| TiN/Si$_3$N$_4$/GZO/SiN (Fig. 3i-b) | 6.13 | 8.4 | 0.55 | 8 | 15 | 0.5 – 1 |
| Si$_3$N$_4$/TiN/Si$_3$N$_4$/GZO/SiN (Fig. 3i-c) | 6.41 | 28 | 12.2 | 16 | 1.3 | 0.2 – 0.3 |
| Si/Si$_3$N$_4$/GZO/Si (Fig. 3ii-a) | 6.79 | 24 | 0.060 | 24 | 400 | 0.09 – 6 |
| TiN/Si$_3$N$_4$/GZO/Si (Fig. 3ii-b) | 7.82 | 60 | 4.2 | 56 | 13 | 0.03 – 0.2 |
| Si/TiN/Si$_3$N$_4$/GZO/Si (Fig. 3ii-c, asym) | 9.00 | 132 | 46 | 86 | 1.9 | 0.04 – 0.09 |
| Si/TiN/Si$_3$N$_4$/GZO/Si (Fig. 3ii-c, sym) | 6.60 | 46 | 0.29 | 46 | 160 | 0.06 – 1.3 |

The extinction ratio for the investigated structures was found to be 1.8-16 dB/μm for a silicon nitride cladding, and 24-86 dB/μm for a silicon cladding. It is one of the highest values reported so far for both theoretical predictions and experimental demonstration for plasmonic modulators.

## 4. Waveguide and modulator integration

Efficient modulators allow a 3 dB modulation depth within a one-micron length plasmonic modulator. Thus, these devices can be very short and considered as a small section of a larger plasmonic waveguide. To couple into these devices, several possible integration schemes can be studied. Modulator structures can be fabricated on top of a plasmonic waveguide. Here we consider 10-nm thin TiN layer, which supports the LR-SPP. In a silicon nitride cladding, the propagation length in such waveguides is 5.5 mm. By limiting the GZO to only a small section required for modulation, the added propagation losses in the GZO dynamic layer are avoided in the remainder of the waveguide. To achieve this, the GZO can either be added directly on top of the TiN layer or used as the plasmonic material in replacement of the TiN (Fig. 5). In the latter case, GZO serves as both a waveguide and dynamic element.

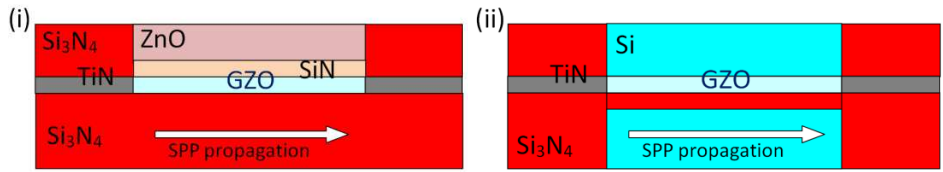

Fig. 5. Schematic of plasmonic modulators integrated with TiN strip waveguides providing long range SPP propagation to and from the modulator (side view). Within the modulator section of the designs, the TiN strip waveguide is replaced by a GZO layer which is plasmonic at the wavelength of interest. The carrier concentration of this GZO film is altered by the application of an electric field across the layer. In geometry (i) the voltage is applied between the GZO and the ZnO layers, (ii) the voltage is applied between the upper and lower silicon layers.

The designs with thick TiN layers (Fig. 3b) could be integrated with a single-interface waveguide. However, such waveguides have significantly higher losses in comparison with thin TiN layers. Therefore, their applications are limited. For this reason, these geometries will not be considered in the following analysis.

Since the mode in the modulator section in Fig. 3c is quasi-asymmetric, it must be excited by the asymmetric mode of a strip-waveguide. Besides the challenge of excitation of asymmetric mode, it has much higher propagation losses. Furthermore, for the design in Fig. 3b,c, a thin layer of $Si_3N_4$ on top or beneath TiN is needed to insulate the GZO layer. However, realization of the designs with $Si/TiN/Si_3N_4/Si$ waveguides encounters an issue. Because of the drastic difference between refractive indices of Si and $Si_3N_4$, it does not support the symmetric mode. It cannot be easily replaced by a Si/TiN/Si waveguide, as the thin $Si_3N_4$ layer (or a more advanced method of electrical isolation such as p-n junction doping) is required to maintain electrical isolation in the active section.

Thus, from the integration point of view, the best modulator structures are those without TiN (Figs. 3a and 5). As we showed in the previous section, these structures also give the highest performance in terms of modulation depth and propagation losses.

Similar to the previous sections, we perform calculations for one-dimensional structures as their properties are close to those of finite-width. The coupling losses γ for a single interface was calculated by following equation

$$\gamma = \frac{4\beta_1\beta_2}{(\beta_1+\beta_2)^2} \frac{\left|\int_\infty E_{1z}E_{2z}^* dz\right|^2}{\int_\infty E_{1z}E_{1z}^* dz \cdot \int_\infty E_{2z}E_{2z}^* dz} ,\qquad(4)$$

where $\beta_1$ ($E_1$) and $\beta_2$ ($E_2$) are the mode indices (electric field) of the waveguide and waveguide modulator, respectively. Eq. (4) it takes into account both the mode overlap integral and the Fresnel coefficients at the boundary region.

We calculated the coupling losses for the two designs shown in Fig. 5, and the results are shown on Fig. 6. With regards to the scheme in Fig. 5i (the silicon nitride cladding), there is an interplay of two major effects. First, coupling losses are increased at the plasmonic resonance because of the high field localization in the GZO layer. This greatly reduces the mode overlap (Fig. 7i). Second, coupling losses are increased at lower and higher carrier concentration in GZO due to the mode extension outside the waveguide where GZO no longer supports a plasmonic mode (similar to Fig. 3 i-f for the geometry "without TiN" where the usable carrier concentration is between approximately 5 and $8\times10^{20}$ cm$^{-3}$). As a result, coupling losses vary by only 5 dB across the modulation range.

For the scheme employing a high-index silicon cladding (Fig. 5ii), the coupling loss in the transmittive state monotonically increases as the carrier concentration decreases towards the maximum in the modulator absorption (similar to the first effect for the low-index cladding). However, because of the mode mismatch at the low-to-high index interface between the waveguide and modulator (Fig. 7i), coupling losses are higher at the GZO plasmonic resonance. This effect can be beneficial for modulator performance in specific applications as it provides additional losses in the resonant state and fewer losses in the transmittive state.

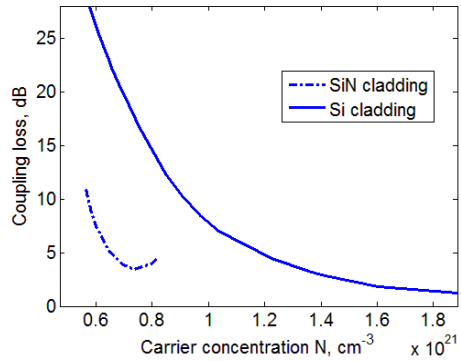

Fig. 6. Single interface coupling loss (between waveguide and modulator sections) versus carrier concentration in the GZO layer. This is shown for both the low-index silicon nitride and high-index silicon cladding. The significant increase of coupling losses corresponds to mode extension outside the waveguide. For the high-index cladding, the coupling losses is large because of the mode size mismatch between the $Si_3N_4/TiN/Si_3N_4$ waveguide and modulator with silicon cladding. For the low-index cladding, we consider only the values of N which correspond to mode size less than 10 μm.

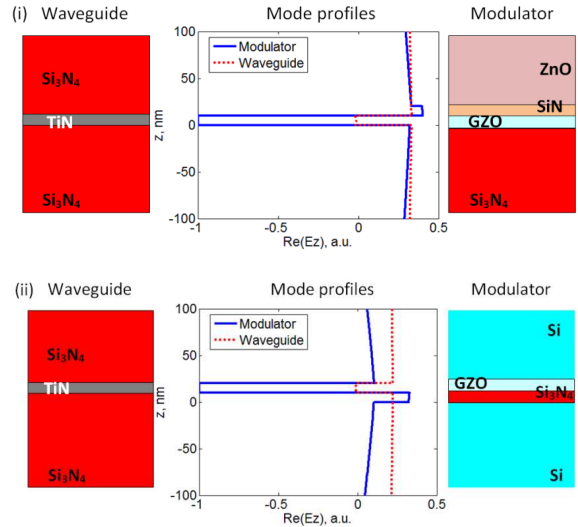

Fig. 7. Example mode profiles in the two integrated modulator geometries: (i) low-index and (ii) high-index claddings. Note that the field decay outside the strip waveguide is slow and therefore appears constant in this graph. The carrier concentration in the GZO layer used for the calculations corresponds to the maximum absorption in the modulator, i.e. plasmonic resonance in the layer. Under these conditions the majority of the field is localized within the GZO layer.

## 5. Conclusion

In this paper, we have analyzed several multilayer structures with alternative plasmonic materials to be utilized in ultra-compact plasmonic modulators. Applying an electric field across the TCO layer allows for the permittivity to be tuned, resulting in a change of the absorption coefficient of the waveguide. Therefore, active modulation is achieved. Numerous modulator layouts are investigated and the typical trade-off between compactness and propagation loss is analyzed. Amongst all the reported structures, one stands out with a remarkable FoM = 400. This figure of merit takes into account both the modulation depth (ER = 24 dB/μm) and the propagation losses in the transmittive state ($\alpha$ = 0.06 dB/μm). The corresponding geometry may allow for ultra-compact modulation with effective length much less than 1 μm. The proposed approach based on the cost-effective planar fabrication processes and the ability to easily integrate with existing semiconductor systems could enable new devices for applications in on-chip optics, sensing, optoelectronics, data storage, and information processing.

**Acknowledgments**

We thank Jieran Fang, Jongbum Kim, and Naresh K. Emani for helpful discussions. We are grateful to Dr. Marcello Ferrera for helping us in preparing this manuscript. V.E.B. acknowledges financial support from Otto Mønsteds Fond and Thriges Fond. A.V.L. acknowledges partial financial support from the Danish Research Council for Technology and Production Sciences via the THz COW project. We acknowledge support from the following grants: ARO grant 57981-PH (W911NF-11-1-0359), NSF MRSEC grant DMR-1120923 and NSF PREM DRM-0611430.

**References**

1. J.A. Dionne and H.A. Atwater, "Plasmonics: Metal-worthy methods and materials in nanophotonics," MRS Bulletin 37, 717-724 (2012).
2. M.L. Brongersma and V.M. Shalaev, "Applied Physics: The case for plasmonics," Science 328, 440 (2010).
3. V.J. Sorger, R.F. Oulton, Ren-Min Ma, and X. Zhang, "Toward integrated plasmonic circuits," MRS Bulletin 37, 728-738 (2012).
4. Surface Plasmon Nanophotonics, Ed. by Mark L. Brongersma and Pieter G. Kik, Springer Netherlands, 2007
5. Plasmonic Nanoguides and Circuits, Ed. by Sergey Bozhevolnyi, Pan Stanford Publishing (2008).
6. Pierre Berini, "Long-range surface plasmon polaritons," Advances in Optics and Photonics 1, 484-588 (2009).
7. A. Boltasseva, T. Nikolajsen, K. Leosson, K. Kjaer, M. S. Larsen, and S. I. Bozhevolnyi, "Integrated optical components utilizing long-range surface plasmon polaritons," J. Lightwave. Technol. 23, 413-422 (2005).
8. Robert Charbonneau, Nancy Lahoud, Greg Mattiussi, and Pierre Berini, "Demonstration of integrated optics elements based on long-ranging surface plasmon polaritons," Opt. Express 13, 977-984 (2005).
9. R. Charbonneau, C. Scales, I. Breukelaar, S. Fafard, N. Lahoud, G. Mattiussi, and P. Berini, "Passive integrated optics elements based on long-range surface plasmon polaritons," J. Lightwave. Technol. 24, 477-494 (2006).

10. V.J. Sorger, Z. Ye, R.F. Oulton, Y. Wang, G. Bartal, X. Yin, and X. Zhang, "Experimental demonstration of low-loss optical waveguiding at deep sub-wavelength scales," Nature Communications 2, 331-1-5 (2011).
11. V.S. Volkov, Z. Han, M.G. Nielsen, K. Leosson, H. Keshmiri, J. Gosciniak, O. Albrektsen, and S. I. Bozhevolnyi, "Long-range dielectric-loaded surface plasmon polariton waveguides operating at telecommunication wavelengths," Optics Letters 36, 4278-4280 (2011).
12. Thomas Nikolajsen, Kristjan Leosson, Sergey I Bozhevolnyi, "Surface plasmon polariton based modulators and switches operating at telecom wavelengths," Applied Physics Letters 85, 5833-5835 (2004).
13. Thomas Nikolajsen, Kristjan Leosson, Sergey I. Bozhevolnyi, "In-line extinction modulator based on long-range surface plasmon polaritons," Optics Communications 244, 455–459 (2005).
14. J.A. Dionne, K. Diest, L.A. Sweatloc, and H.A. Atwater, "PlasMOStor: a metal- oxide- Si field effect plasmonic modulator," Nano Letters 9, 897-902 (2009).
15. W. Cai, J. S. White, and M.L. Brongersma, "Compact, high-speed and power-efficient electrooptic plasmonic modulators," Nano Letters 9, 4403 (2009).
16. K.F. MacDonald, N.I. Zheludev, "Active plasmonics: current status," Laser Photon. Rev. 4, 562 (2010).
17. A. Melikyan, N. Lindenmann, S. Walheim, P.M. Leufke, S. Ulrich, J. Ye, P. Vincze, H. Hahn, Th. Schimmel, C. Koos, W. Freude, and J. Leuthold, "Surface plasmon polariton absorption modulator," Optics Express 10, 8855-8869 (2011).
18. A.V. Krasavin and A.V. Zayats, "Photonic Signal Processing on Electronic Scales: Electro-Optical Field-Effect Nanoplasmonic Modulator," Phys. Rev. Lett. 109, 053901-1-5 (2012).
19. V.J. Sorger, N.D. Lanzillotti-Kimura, R.-M. Ma, and X. Zhang, "Ultra-compact silicon nanophotonic modulator with broadband response," Nanophotonics 1, 17-23 (2012).
20. Luke A. Sweatlock and Kenneth Diest, "Vanadium dioxide based plasmonic modulators," Opt. Express 20, 8700-8709 (2012).
21. V.E. Babicheva, I.V. Kulkova, R. Malureanu, K. Yvind, A.V. Lavrinenko, "Plasmonic modulator based on gain-assisted metal–semiconductor–metal waveguide," Photonics and Nanostructures-Fundamentals and Applications 10, 389 (2012).
22. A. Melikyan, L. Alloatti, A. Muslija, D. Hillerkuss, P. Schindler, J. Li, R. Palmer, D. Korn, S. Muehlbrandt, D. Van Thourhout, B. Chen, R. Dinu, M. Sommer, C. Koos, M. Kohl, W. Freude, and J. Leuthold, "Surface Plasmon Polariton High-Speed Modulator," in CLEO: 2013, OSA Technical Digest, paper CTh5D.2 (2013).
23. A. Boltasseva and H.A. Atwater, "Low-loss plasmonic metamaterials," Science 331, 290-291 (2011).
24. P.R. West, S. Ishii, G. Naik, N. Emani, V.M. Shalaev, and A. Boltasseva, "Searching for better plasmonic materials," Laser & Photonics Reviews 4, 795-808 (2010).
25. Crissy Rhodes, Stefan Franzen, Jon-Paul Maria, Mark Losego, Donovan N. Leonard, Brian Laughlin, Gerd Duscher, and Stephen Weibel, "Surface plasmon resonance in conducting metal oxides", J. Appl. Phys. 100, 054905 (2006).
26. G.V. Naik and A. Boltasseva, "Semiconductors for plasmonics and metamaterials," Phys. Status Solidi RRL 4, 295-297 (2010).
27. G.V. Naik and A. Boltasseva, "A comparative study of semiconductor-based plasmonic metamaterials," Metamaterials 5, 1-7 (2011).
28. G.V. Naik, J. Kim, and A. Boltasseva, "Oxides and nitrides as alternative plasmonic materials in the optical range," Optical Materials Express 1, 1099 (2011).
29. G.V. Naik, J.L. Schroeder, X. Ni, A.V. Kildishev, T.D. Sands, and A. Boltasseva, "Titanium nitride as plasmonic material for visible wavelengths," Optical Materials Express 2, 478-489 (2012).
30. G.V. Naik, J. Liu, A.V. Kildishev, V.M. Shalaev, and A. Boltasseva, "A demonstration of Al:ZnO as a plasmonic component for near-infrared metamaterials," PNAS 109, 8834-8838 (2012).
31. J.B. Khurgin and A. Boltasseva, "Reflecting upon losses in plasmonics and metamaterials," MRS Bull. 3, 768-779 (2012).
32. G. Naik, V.M. Shalaev, and A. Boltasseva, "Alternative plasmonic materials: beyond gold and silver," Advanced Materials 25, 3264–3294 (2013).
33. J. Narayan, P. Tiwari, X. Chen, J. Singh, R. Chowdhury, and T. Zheleva, "Epitaxial growth of TiN films on (100) silicon substrates by laser physical vapor deposition," Appl. Phys. Lett. 61, 1290 (1992).
34. E. Feigenbaum, K. Diest, and H.A. Atwater, "Unity-order index change in transparent conducting oxides at visible frequencies," Nano Letters 10, 2111-2116 (2010).
35. Z. Lu, W. Zhao, and K. Shi, "Ultracompact Electroabsorption Modulators Based on Tunable Epsilon-Near-Zero-Slot Waveguides," IEEE Photonics Journal 4, 735-740 (2012).
36. Kazuto Noguchi, Osamu Mitomi, and Hiroshi Miyazawa, "Millimeter-Wave Ti:LiNbO3 Optical Modulators," J. Lightwave Technol. 16, 615 (1998).
37. V. E. Babicheva and A. Lavrinenko, "Plasmonic modulator optimized by patterning of active layer and tuning permittivity," Optics Communications 285, 5500-5507 (2012).
38. B. Lamprecht, J.R. Krenn, G. Schider, H. Ditlbacher, M. Salerno, N. Felidj, A. Leitner and F.R. Aussenegg, J. C. Weeber, "Surface plasmon propagation in microscale metal stripes," Appl. Phys. Lett. 79, 51–53 (2001).
39. J. Kim, G.V. Naik, N.K. Emani, U. Guler, and A. Boltasseva, "Plasmonic resonances in nanostructured transparent conducting oxide films," IEEE Journal of Selected Topics in Quantum Electronics 19, 4601907 (2013).
40. M. Bass, C. DeCusatis, G. Li, V. N. Mahajan, E. V. Stryland, Handbook of Optics, Volume II: Design, fabrication and testing, sources and detectors, radiometry and photometry, McGraw-Hill: New York, 1994.
41. Sopra data sheet, http://www.sspectra.com/sopra.html
42. V.E. Babicheva, R. Malureanu, A.V. Lavrinenko, "Plasmonic finite-thickness metal-semiconductor-metal waveguide as ultra-compact modulator," Photonics and Nanostructures-Fundamentals and Applications accepted, http://arxiv.org/abs/1301.5603